\documentclass[11pt]{article}

\usepackage{fullpage,amsmath,amsthm,amssymb,amscd,fancybox,epsfig,float,subfigure}
\usepackage{graphics}

\usepackage{amsmath, url}
\usepackage{ifpdf}
 \ifpdf
 \usepackage[numbers]{natbib}
\fi

\newcommand{\beq} {\begin{equation}}
\newcommand{\eeq} {\end{equation}}
\newcommand{\bit}{\begin{itemize}}
\newcommand{\eit}{\end{itemize}}
\newcommand{\bde}{\begin{description}}
\newcommand{\ede}{\end{description}}
\newcommand{\bce}{\begin{center}}
\newcommand{\ece}{\end{center}}
\newcommand{\ben} {\begin{enumerate}}
\newcommand{\een} {\end{enumerate}}
\newcommand{\bea} {\begin{eqnarray}}
\newcommand{\eea} {\end{eqnarray}}
\newcommand{\barr} {\begin{array}}
\newcommand{\earr} {\end{array}}
\newcommand{\bean} {\begin{eqnarray*}}
\newcommand{\eean} {\end{eqnarray*}}

\newcommand{\sgl}{\mathcal{S}}
\newcommand{\mcd}{\mathcal{D}}

\newcommand{\mbe}{\mathbf{E}}
\newcommand{\mbx}{\mathbf{x}}
\newcommand{\mby}{\mathbf{y}}

\newcommand{\mbn}{\mathbf{n}}

\newcommand{\mbf}{\mathbf{f}}

\newcommand{\norm}[1]{|\!| #1 |\!|}
\newcommand{\mbv}{\mathbf{v}}

\newcommand{\jump}[1]{\left[\!\left[ #1 \right]\!\right]_\gamma}

\usepackage{color}
\definecolor{myblue}{rgb}{.85, .85, 1}
\usepackage{empheq}
\newlength\mytemplen
\newsavebox\mytempbox

\makeatletter
\newcommand\mybluebox{%
    \@ifnextchar[
       {\@mybluebox}%
       {\@mybluebox[0pt]}}

\def\@mybluebox[#1]{%
    \@ifnextchar[
       {\@@mybluebox[#1]}%
       {\@@mybluebox[#1][0pt]}}

\def\@@mybluebox[#1][#2]#3{
    \sbox\mytempbox{#3}%
    \mytemplen\ht\mytempbox
    \advance\mytemplen #1\relax
    \ht\mytempbox\mytemplen
    \mytemplen\dp\mytempbox
    \advance\mytemplen #2\relax
    \dp\mytempbox\mytemplen
    \colorbox{myblue}{\hspace{1em}\usebox{\mytempbox}\hspace{1em}}}

\makeatother

\newcommand{\bemp}{\begin{empheq}[box={\mybluebox[5pt]}]{equation}}
\newcommand{\eemp}{\end{empheq}}

\title{Electrohydrodynamics of deflated vesicles: budding, rheology and pairwise interactions}

\author{Bowei Wu \, and \, Shravan Veerapaneni\thanks{Department of Mathematics, University of Michigan. Authors gratefully acknowledge support from NSF under grants DMS-1418964 and DMS-1454010. 
}
}

\begin{document}

\maketitle
\begin{abstract}
The electrohydrodynamics of vesicle suspensions is characterized by studying their pairwise interactions in applied DC electric fields in two dimensions. In the dilute limit, the rheology of the suspension is shown to vary nonlinearly with the electric conductivity ratio of the interior and exterior fluids. The prolate-oblate-prolate transition and other transitionary dynamics observed in experiments and previously confirmed via numerical simulations is further investigated here for smaller reduced areas. When two vesicles are initially un-aligned with the external electric field, three different responses are observed when the key parameters are varied: (i) {\em chain formation}--they self-assemble to form a chain that is  aligned along the field direction, (ii) {\em circulatory motion}--they rotate about each other, (iii) {\em oscillatory motion}--they form a chain but oscillate about each other. 
\end{abstract}
\maketitle

\section{Introduction}

Understanding the electrohydrodynamics (EHD) of the so-called giant unilamellar vesicles (GUVs) has received much attention in the recent past \citep{perrier2017lipid}. Vesicles share the same structural component of a biological cell, the bilipid membrane, and hence their EHD has been a paradigm for understanding how general biological cells behave under an electric field. The dynamics of this system is characterized by a competition between viscous, elastic, and electric stresses on the individual membranes and the nonlocal hydrodynamic interactions. Studying the microstructural response of isolated vesicles and vesicle pairs subjected to electric fields can bring insights into the macroscopic properties of vesicle suspensions. Several recent theoretical and numerical works have focused on the former case but to our knowledge, detailed analysis of the latter is lacking.  In this work, we characterize, through numerical simulations,  the pairwise hydrodynamics of vesicles subjected to a uniform DC electric field.

Theoretical investigation of vesicle EHD has been done via small deformation theory \citep{vlahovska2009electrohydrodynamic, schwalbe2011vesicle} and semi-analytic studies using spheroidal models \citep{zhang2013transient, nganguia2013}. Numerical solution of the coupled electric, elastic and hydrodynamic governing equations were computed using the boundary integral equation (BIE) methods \citep{mcconnell2013vesicle, salipante2014vesicle, ehd3d} and immersed interface or immersed boundary methods \citep{kolahdouz2015dynamics, hu2016vesicle}. Advantages of BIE methods are well-known---exact satisfaction of far-field boundary conditions eliminating the need for artificial boundary conditions, reduction in dimensionality leading to reduced problem sizes, and well-conditioned linear systems through carefully chosen integral representations.  

All of the aforementioned works, however, considered EHD of a single vesicle only. Vesicles are known to segregate when subjected to electric fields \citep{ristenpart2010dynamic}, thereby, pose significant challenges for direct numerical simulations. In the case of BIE methods, for instance, the integral representations of the hydrodynamic and electric interaction forces become nearly-singular, requiring specialized quadratures.  Domain discretization methods, on the other hand, require finer meshes (locally, in the case of adaptive methods), worsening the conditioning issue of linear systems and increasing the overall computational expense. 

Leveraging on our recently developed spectrally-accurate algorithm for evaluating nearly singular integrals \citep{lsc2d} and the second-kind BIE formulation for three-dimensional vesicle EHD \citep{ehd3d}, we develop a BIE method for simulating multiple vesicle EHD in this work. We apply it to analyze the pairwise interactions in a monodisperse suspension. We provide the integral equation formulation and the description of our numerical method in \S\ref{sc:formulate}, followed by analysis and discussion of the results in \S\ref{sc:results}.   

\section{Problem formulation} \label{sc:formulate}

\subsection{Governing equations}
Let us first consider a single vesicle suspended in a two-dimensional unbounded viscous fluid domain, subjected to an imposed flow $\mbv_\infty(\mbx)$, for any $\mbx \in \mathbb{R}^2$. The vesicle membrane is denoted by $\gamma$. Assume that the fluids interior and exterior to $\gamma$ have the same viscosity $\mu$ and the same dielectric permittivity $\epsilon$ while their conductivities differ, given by $\sigma_i$ and $\sigma_e$, respectively. In the vanishing Reynolds number limit, the governing equations for the ambient fluid can then be written as: 
\begin{subequations}
\begin{align}
-\nabla p + \mu \triangle \mbv = 0 \quad\text{in}\quad \mathbb{R}^2\setminus\gamma,  &  \\
\nabla \cdot \mbv = 0  \quad\text{in}\quad \mathbb{R}^2\setminus\gamma, &\\
\mbv(\mbx) \rightarrow \mbv_\infty(\mbx) \quad \text{as} \quad  \norm{\mbx} \rightarrow \infty. & 
 \end{align} \label{eqn:governing}
\end{subequations}%
The fluid motion is coupled to the membrane motion via the kinematic boundary condition $\dot{\mbx} = \mbv$ on $\gamma$,
where $\mbx$ is a material point on the membrane. Using the boundary integral equation formulation, we can now write the membrane evolution equation by combining the kinematic condition with the governing equation \eqref{eqn:governing} as \citep{ves2d},
\begin{empheq}[box={\mybluebox[5pt]}]{equation}
\dot{\mbx} = \mbv_\infty (\mbx) + \int_\gamma G_s(\mbx - \mby) \mbf_{hd}(\mby) \, d\gamma (\mby), \quad \nabla_\gamma \cdot \dot{\mbx} = 0, \label{eq:IE_HI}
\end{empheq}
where $\mbf_{hd}$ is the hydrodynamic traction jump across the membrane and $G_s$ is the free-space Green's function for the Stokes equations, given by 
\beq
      G_s(\mbx - \mby) = \frac{1}{4\pi\mu}\left(-\log \norm{\mbx - \mby}\,
       \mathbf{I} + \frac{(\mbx - \mby) \otimes (\mbx - \mby)}{\norm{\mbx - \mby}^2}
         \right).
    \label{eqn:Gs}
 \eeq
The second equation in (\ref{eq:IE_HI}) expresses the local inextensibility constraint on the membrane. 

For a given vesicle configuration, $\mbf_{hd}$ can be evaluated by performing a force balance at the membrane. The elastic forces acting on the membrane are comprised of the bending and the tension forces, defined respectively as 
\beq\mbf_b = \kappa_B \, \left( \kappa_{ss} + \frac{\kappa^3}{2}\right)\,\mbn, \quad \mbf_\lambda = (\lambda \mbx_s)_s, 
 \eeq
where $\kappa_B$ is the bending modulus, $\kappa$ is the curvature, $s$ is the arclength parameter, $\mbn$ is the outward normal to $\gamma$ and the tension $\lambda$ acts as a Lagrange multiplier to enforce the inextensibility constraint. A force balance at the membrane yields $\mbf_{hd} = \mbf_b + \mbf_\lambda - \mbf_{el}$, where $\mbf_{el}$ is the electric force that is determined by solving for the electric potential.

In the {\em leaky--dielectric} model, the electric charges are assumed to be present only at the interface and not in the bulk. Let $\phi(\mbx)$ be the electric potential at $\mbx$, so that $\mbe = -\nabla \phi$. Assuming that the vesicle membrane is charge-free and has a conductivity $G_m$, a capacitance $C_m$, the boundary value problem for the electric potential can be summarized as \citep{schwalbe2011vesicle}:  
\begin{subequations}
\begin{align}
-\triangle \phi = 0   \quad\text{in}\quad \mathbb{R}^2\setminus\gamma, & \\
-\nabla \phi(\mbx) \rightarrow \mbe_\infty(\mbx) \quad \text{as} \quad \norm{\mbx} \rightarrow \infty, \qquad \jump{\mbn\cdot (\sigma \nabla \phi)} = 0,  \qquad \jump{\phi} = V_m, & \label{eq:mem_potential}\\
C_m\dot{V}_m + G_mV_m = -\mbn \cdot (\sigma_i\nabla \phi_i) \quad \text{on} \quad \gamma. & \label{eq:current_conserv}
\end{align} \label{eqn:potential}
\end{subequations}
Here, $\mbe_\infty$ is the imposed electric field, $\jump{\cdot}$ denotes the jump across the interface (e.g., $\jump{\sigma} = \sigma_i - \sigma_e$) and  $V_m$ is the transmembrane potential. The electric force on the membrane is then defined by  
$\mbf_{el} = \jump{\mbn\cdot\Sigma^{el}}$, where the Maxwell stress tensor,  $\Sigma^{el} =  \epsilon \mbe \otimes \mbe - \frac{1}{2}\epsilon \norm{\mbe}^2 \,  \mathbf{I}$. 
Therefore, we need to determine the electric field on both sides of the membrane by solving (\ref{eqn:potential}) to evaluate $\mbf_{el}$.

Since we are only interested in interfacial variables and (\ref{eqn:potential}) is a linear partial differential equation, we can recast it as a BIE with the unknowns residing only on the interface. We will employ an {\em indirect} integral equation formulation to solve for the electric potential $\phi$. Assume that the electric potential in the domain interior and exterior of the membrane is given by \citep{ehd3d},
\begin{empheq}[box={\mybluebox[5pt]}]{equation}
\phi(\mbx) =\phi_\infty(\mbx) + \sgl [q](\mbx) -  \mcd[V_m](\mbx) \label{potential}
\end{empheq}
where the membrane charge density, $q = \jump{\partial\phi/\partial \mbn}$ and the Laplace single and double layer integral operators are defined by 
\beq \sgl[q](\mbx) = \int_\gamma G(\mbx-\mby) q(\mby) \, d\gamma(\mby) \quad \text{and} \quad 
 \mcd[V_m](\mbx) = \int_\gamma \dfrac{\partial G(\mbx - \mby)}{\partial \mbn(\mby)} V_m(\mby) \, d\gamma(\mby), \label{eq:single_layer}\eeq respectively. Here $G(\cdot)$ is the Laplace fundamental solution in the free space.

Note that, by construction, equation (\ref{potential}) implies $\jump{\phi} = V_m\,$ since the single layer potential is continuous across $\gamma$. Applying the current continuity condition and using the standard jump conditions for the Laplace layer potentials, we arrive at the {\em second-kind} integral equation for the unknown $q$: 
\begin{empheq}[box={\mybluebox[5pt]}]{equation}
\left(\frac{1}{2} + \eta\, \sgl'\,\right) q= \eta \, \mbe_\infty \cdot \mbn  + \eta\, \mcd' [V_m], \label{q}
\end{empheq}
where $\eta = (\sigma_i - \sigma_e)/(\sigma_i + \sigma_e)$, $\sgl'$ and $\mcd'$ denote the normal derivatives of the single and double layer potentials respectively. Furthermore, the interfacial conditions $\jump{\partial\phi/\partial \mbn} = q$ and  $\jump{\sigma\partial\phi/\partial \mbn} = 0$ imply that $-\mbn \cdot (\sigma_i\nabla \phi_i) = (\sigma_i\sigma_e/(\sigma_i-\sigma_e))q$. Substituting this result in (\ref{eq:current_conserv}) and using (\ref{q}), we arrive at the following integro-differential equation for the evolution of $V_m$:
\begin{empheq}[box={\mybluebox[5pt]}]{equation}
C_m\dot{V}_m + G_mV_m =  \dfrac{\sigma_i\sigma_e}{\sigma_i+\sigma_e}\left(\frac{1}{2} + \eta\, \sgl'\,\right)^{-1}( \mbe_\infty \cdot \mbn  + \mcd' [V_m]). \label{Vm}
\end{empheq}

The steps involved within a time-stepping procedure for the electric problem for a given vesicle shape can now be summarized as follows: update $V_m$ using \eqref{Vm}, which also gives $q$ since the right-hand side of \eqref{Vm} is just $(\sigma_i\sigma_e/(\sigma_i-\sigma_e))q$, then evaluate the membrane electric force $\mbf_{el}$ by computing $\mbe_i$ and $\mbe_e$ using \eqref{potential}. 

Finally, the formulation generalizes to the two- (or multiple-) vesicle case in a trivial manner. Let $\gamma$ now denote the union of the vesicle membranes i.e., $\gamma = \bigcup_{i=1}^2 \gamma_i $, where $\gamma_i$ is the boundary of the $i$-th vesicle. Then, the definition of the boundary integral operators introduced earlier hold as is; for example, 
\beq
\sgl[q](\mbx) = \int_\gamma G(\mbx-\mby) q(\mby) \, d\gamma(\mby) := \sum_{j = 1}^{2} \int_{\gamma_j} G(\mbx-\mby) q(\mby) \, d\gamma_j(\mby).
\eeq

\subsection{Numerical Method}\label{sc:numerical}
We now describe a numerical scheme to solve the coupled integro-differential equations for the evolution of vesicle position \eqref{eq:IE_HI} and its transmembrane potential \eqref{Vm}. It directly follows from ideas introduced in \cite{ves2d}, \cite{lsc2d} and \cite{ehd3d}. Each vesicle boundary is parametrized by a Lagrangian variable $\alpha \in [0, 2\pi]$ and a uniform discretization in $\alpha$ is employed. Derivatives of functions defined on the boundary are then computed using spectral differentiation in the Fourier domain, accelerated by the fast Fourier transform.  

\noindent {\em \textbf{Evaluating boundary integrals. }} 
We use the standard periodic trapezoidal rule for computing boundary integrals that are smooth (e.g., the double-layer potential defined in \eqref{eq:single_layer}), which yields spectral accuracy. On the other hand, we discretize the weakly singular operators such as the single-layer potential defined in \eqref{eq:single_layer} using a spectrally-accurate Nystr\"{o}m method (with periodic Kress corrections for the log singularity, (\citet{kress1999linear}, Sec. 12.3)). The same method is also applied for computing the Stokes single-layer potential \eqref{eq:IE_HI}.  

The operator $\mathcal{D}'[\cdot]$ requires special attention as its kernel is {\em hyper-singular}. We employ the following standard transformation (\cite{hsiao2008boundary}) to turn it into a weakly singular integral:
\begin{equation}
\mcd'[V_m](\mbx) = \frac{\partial}{\partial \mbn(\mbx)}\int_\gamma \frac{\partial G(\mbx - \mby)}{\partial \mbn(\mby)} V_m(\mby) \, d\gamma = \frac{\partial}{ \partial s(\mbx)}\left(\int_\gamma G(\mbx - \mby) \frac{\partial V_m(\mby)}{\partial s(\mby)} \, d\gamma \right) \, \forall \, \mbx\in\gamma.
\label{eq:dlp_normal2}
\end{equation}
The surface gradients, $\partial/\partial s(\mbx)$ and $\partial/\partial s(\mby)$, are computed via spectral differentiation. 

Lastly, when the vesicles are located arbitrarily close to each other, the boundary integrals evaluating the interaction forces becomes {\em nearly-singular}. For example, consider the integral, 
\beq \int_{\gamma_1} G(\mbx - \mby) q(\mby) \, d\gamma_1 (\mby), \quad\text{where}\quad \mbx\in\gamma_2. \eeq
The periodic trapezoidal rule loses its uniform spectral convergence in evaluating this integral as $\mbx$ approaches $\gamma_1$; moreover, the singular quadrature rule is also ineffective for this integral. These inaccuracies, in turn, may lead to numerical instabilities and breakdown of the simulation. To remedy this problem, we employ the recently developed close evaluation scheme of \cite{lsc2d} whenever vesicles are located closer than a cutoff distance (which is heuristically chosen to be five times the minimum spacing between the nodes, the so-called ``$5h$-rule''). This scheme achieves spectral accuracy in evaluating \eqref{eq:dlp_normal2}, regardless of the distance of $\mbx$ from $\gamma_1$. We use this scheme to accurately evaluate the Stokes layer potential in (\ref{eq:IE_HI}) as well.

\noindent {\em \textbf{Time-stepping scheme. }} The numerical stiffness associated with the bending force on the vesicle membranes is overcome by using the semi-implicit scheme proposed in \cite{ves2d} to discretize \eqref{eq:IE_HI} in time. Following \cite{mcconnell2013vesicle} and \cite{ehd3d}, we treat the electric force on the membrane explicitly, thereby, decoupling the evolution equations \eqref{eq:IE_HI} and \eqref{Vm}. Then, we use a semi-implicit scheme to evolve the transmembrane potential independently, which we describe next.   

Let $\Delta t$ be the time-step size, $V_m^n(\mbx)$ be the transmembrane potential at time $n\Delta t$ at a point $\mbx$ on the membrane. Our semi-implicit time-stepping scheme for \eqref{Vm} is given by
\begin{equation}
C_{m}\dfrac{V_{m}^{n+1} - V_{m}^{n}}{\Delta t}+ G_{m}V_{m}^{n+1} =  \dfrac{\sigma_i\sigma_e}{\sigma_i+\sigma_e} \left( \dfrac{1}{2} + \eta\,\sgl'\right)^{-1}(\mbe_{\infty}\cdot \mbn + \mcd'V_{m}^{n+1}),
\end{equation}
where the boundary integral operators are treated explicitly i.e., evaluated using the boundary position at $n \Delta t$. This linear system for the unknown $V_m^{n+1}$ is solved using an iterative method (GMRES).

\section{Results and discussions}\label{sc:results}

We now turn to analyzing the simulation results obtained using the numerical method outlined above. We first compare our results on single vesicle EHD with those obtained in prior studies as well as present some new insights on dynamics and rheology of dilute suspensions, followed by analysis of pairwise dynamics. Let $A$ and $L$ denote the area and perimeter of the vesicle. Setting the characteristic length scale as $a =  L/2\pi$, we characterize our results on the following four nondimensional parameters, 

\emph{\begin{tabbing}
bbbbbbbbbbbbbbbbbbbbbbbbbbbbbbbbbbbb\= \kill 
reduced area: \> $\Delta = 4\pi A/L^2$,\\ 
conductivity ratio: \> $ \Lambda =  \sigma_i/\sigma_e$,\\ 
membrane conductivity: \> $G = aG_m/\sigma_{e}$,\\ 
electric field strength: \> $\beta = \epsilon |\mbe_\infty|^2aC_m/\mu\sigma_{e}$, \\
capillary number: \> $\mathit{Ca} =  \mu\dot{\gamma}a^3/\kappa_B$,\\
bending rigidity: \> $\chi = C_m\kappa_B/\sigma_e\mu a^2 $, \\
\end{tabbing}}

\noindent where $\dot{\gamma}$ is the shear rate e.g., for imposed linear shear flow, we have $\mbv_\infty(\mbx) = (\dot{\gamma} x_2, 0)$. In all the simulations, the time is non-dimensionalized by the bending relaxation timescale $t_{\kappa_B} = \mu a^3/\kappa_B$ and the bending rigidity, $\chi \approx 0.08$.
\subsection{Isolated vesicle EHD: transition from squaring to budding in POP} \label{sc:one-ves-deform}

\begin{figure}[t]
\begin{center}
\includegraphics[width=\textwidth]{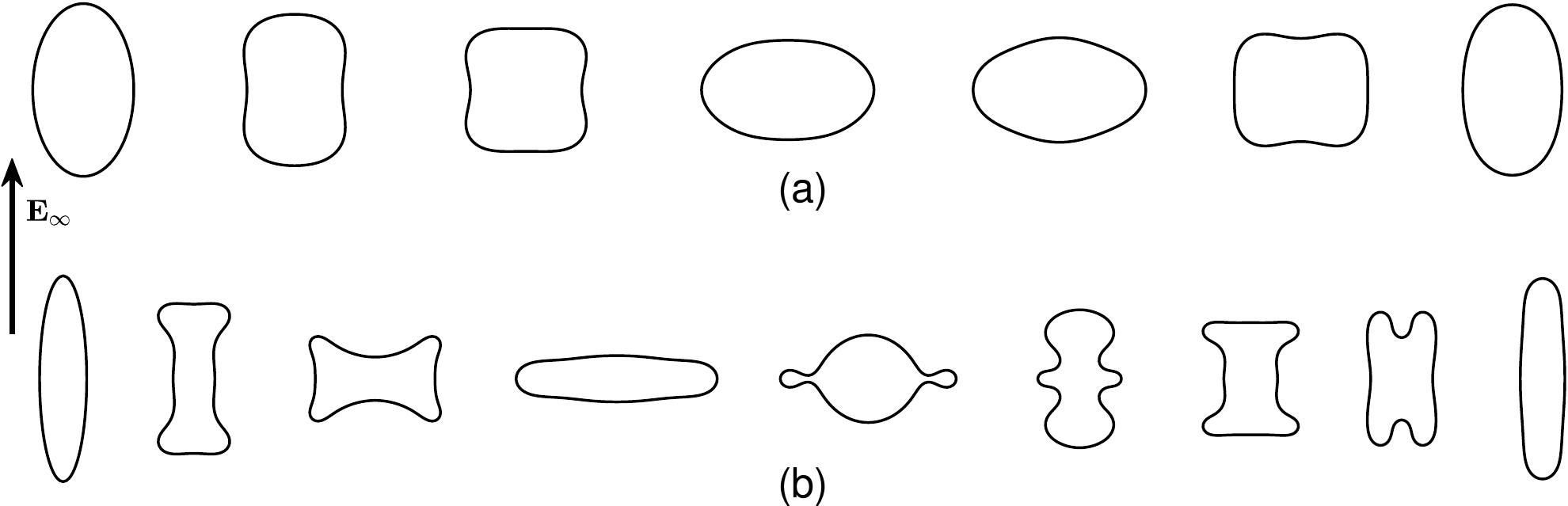}
\caption{Snapshots from two different simulations of a single vesicle subjected to an external electric field, with $\Lambda = 0.1$, $G=0$, $\mathit{Ca} = 0$ and (a) $\Delta = 0.9$, $\beta = 3.2$ and (b) $\Delta = 0.5$, $\beta = 12.8$. While the vesicle undergoes a prolate-oblate-prolate transition, the transient ``square-like'' shapes observed here (in (a)) and in prior numerical studies cannot be observed when the reduced area is lowered. Instead, to sustain the electric compression forces, the vesicle forms buds as it undergoes the POP transition (more details on this phase are shown in Figure \ref{fig:POP5_streamline}).}
\label{fig:one_ves_deform}
\end{center}
\end{figure}

When an arbitrarily shaped vesicle is subjected to uniform electric field, it is known to transform into either a prolate shape or an oblate shape at equilibrium \citep{riske2005electro, Sadik2011}. Since ours is a 2D construct, we refer to ellipses whose major axis aligns with the electric field direction as ``prolates''; similarly, those whose minor axis aligns as ``oblates''. A classical observation in vesicle EHD studies is the {\em prolate-oblate-prolate} (POP) transition that arises in certain parameter regimes. Figure \ref{fig:one_ves_deform}(a) illustrates the POP transition simulated using our numerical method. 

Three conditions are generally required for a vesicle to undergo POP transition: 1) $G$ is very small so that the vesicle membrane acts more like a capacitor than a conductor, 2) $\Lambda$ is less than one and 3) $\beta$ is strong enough. Since $\Lambda < 1$, charges accumulate faster on the membrane exterior initially, thereby, the vesicle appears to be negatively charged at the top and positively charged at the bottom, leading to a compressional force from the applied electric field and the vesicle transitions from a prolate to an oblate shape. At longer times, once the membrane, acting as a capacitor, is fully charged, the apparent charge becomes zero and the vesicle transforms back into a prolate shape, which minimizes the electrostatic energy \citep{C5SM00585J}.  

\begin{figure}[t]
\begin{center}
\includegraphics[height = 1.4in]{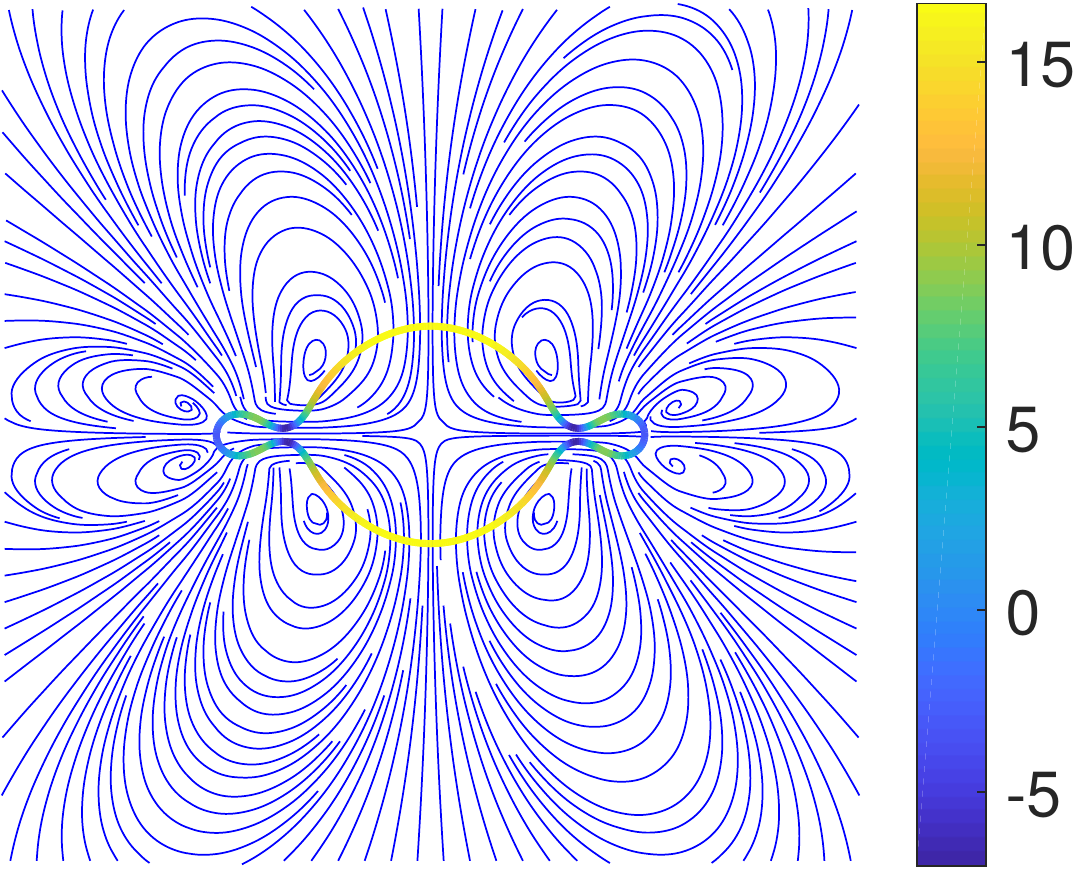} \includegraphics[height = 1.4in]{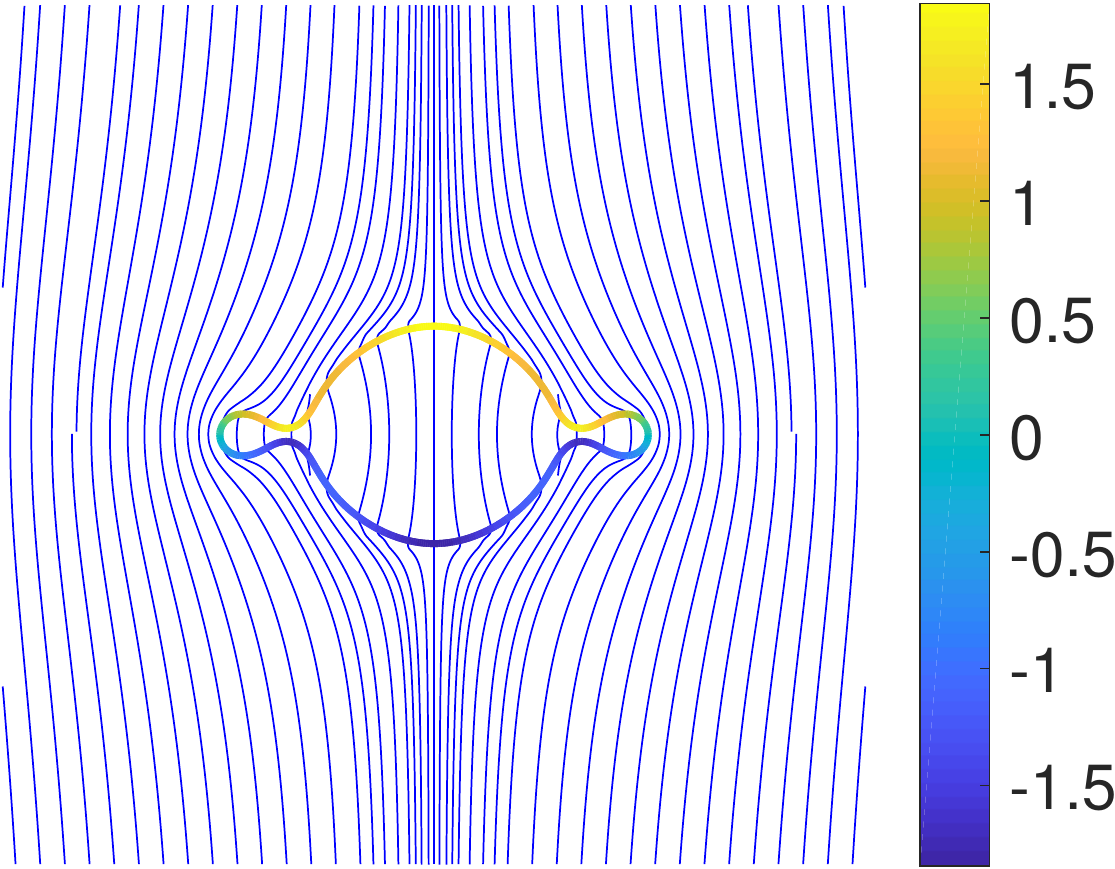}\includegraphics[height = 1.35in]{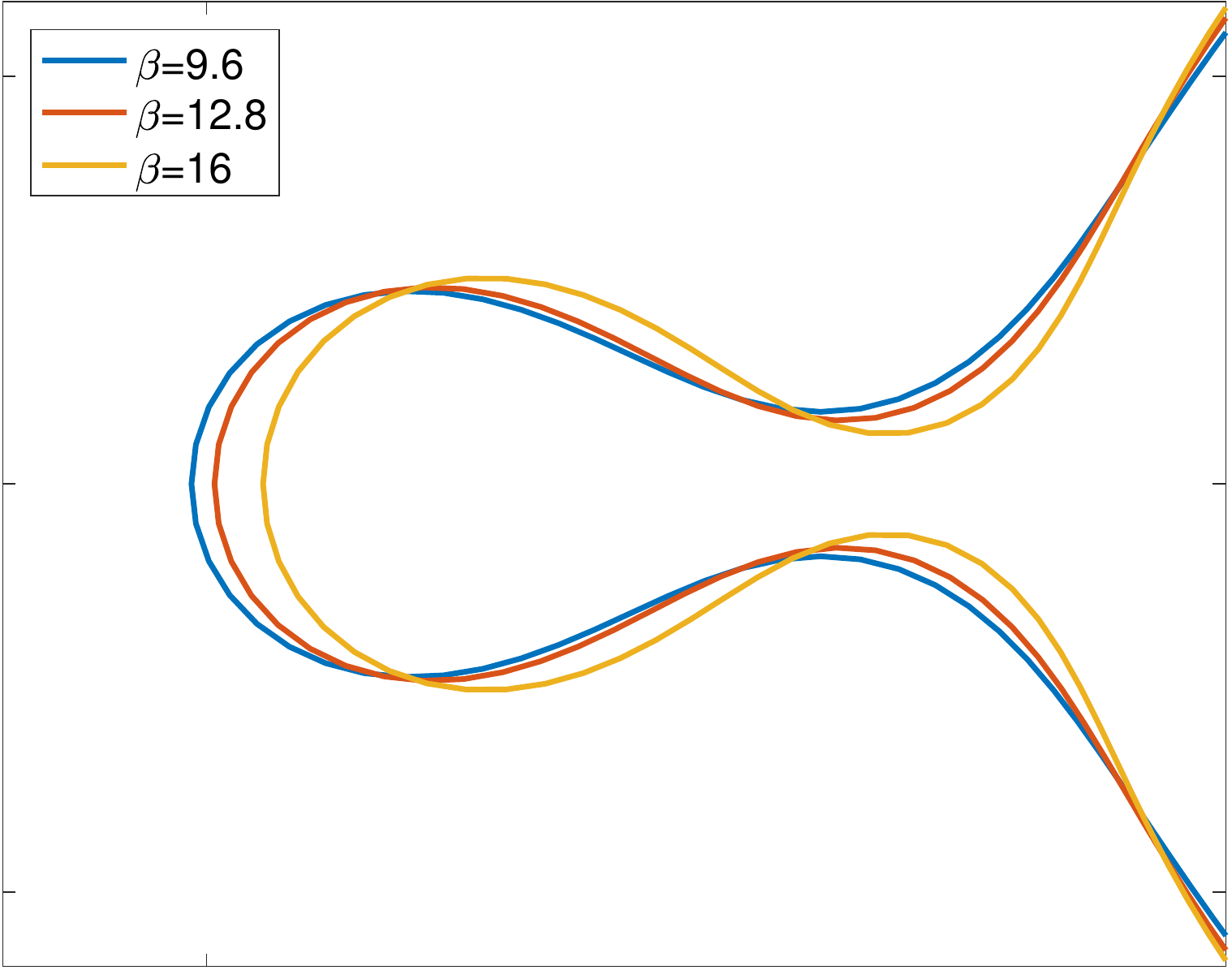}
\caption{Streamlines (left) and electric field lines (middle) plotted at the moment when the vesicle with $\Delta = 0.5$ shown in Figure \ref{fig:one_ves_deform}(b) forms buds while undergoing POP transition. In the left figure, the membrane color indicates the magnitude of tension while on the middle figure, it indicates the magnitude of the transmembrane potential. The right figure gives a closer look at the \emph{narrowest} buds formed under different $\beta$'s, where the times correspond to this state for $\beta = 9.6, 12.8$ and $16$ are $t = 0.253, 0.216$ and $0.184$, respectively. The neck of the buds becomes narrower as $\beta$ increases.}
\label{fig:POP5_streamline}
\end{center}
\end{figure}

A notable feature of the POP transition is the \emph{squaring effect}---a transient shape of the vesicle with four smoothed corners (as can be observed in Figure \ref{fig:one_ves_deform}(a))---which attracted attention of researchers due to its implications on electroporation. Since the reduced area of a square is around 0.785, a question naturally arises:  What transient shapes would a vesicle with much lower reduced area assume? In Figure \ref{fig:one_ves_deform}(b), we illustrate the POP transition of a vesicle with $\Delta = 0.5$. Since the fluid incompressibility acts to preserve its enclosed area, the vesicle forms small protrusions or ``buds'' to sustain the electrical compression forces. Figure \ref{fig:POP5_streamline} shows more details of this bud formation phase. The tension becomes negative, as expected, in the neck region of the buds. These intermediary shapes are reminiscent of those obtained by growing microtubules within the vesicles \citep{fygenson1997mechanics}; the notable feature here, however, is that only body forces are applied as opposed to local microtubule-membrane forces.

\begin{figure}[t]
\begin{center}
\includegraphics[height = 1.4in]{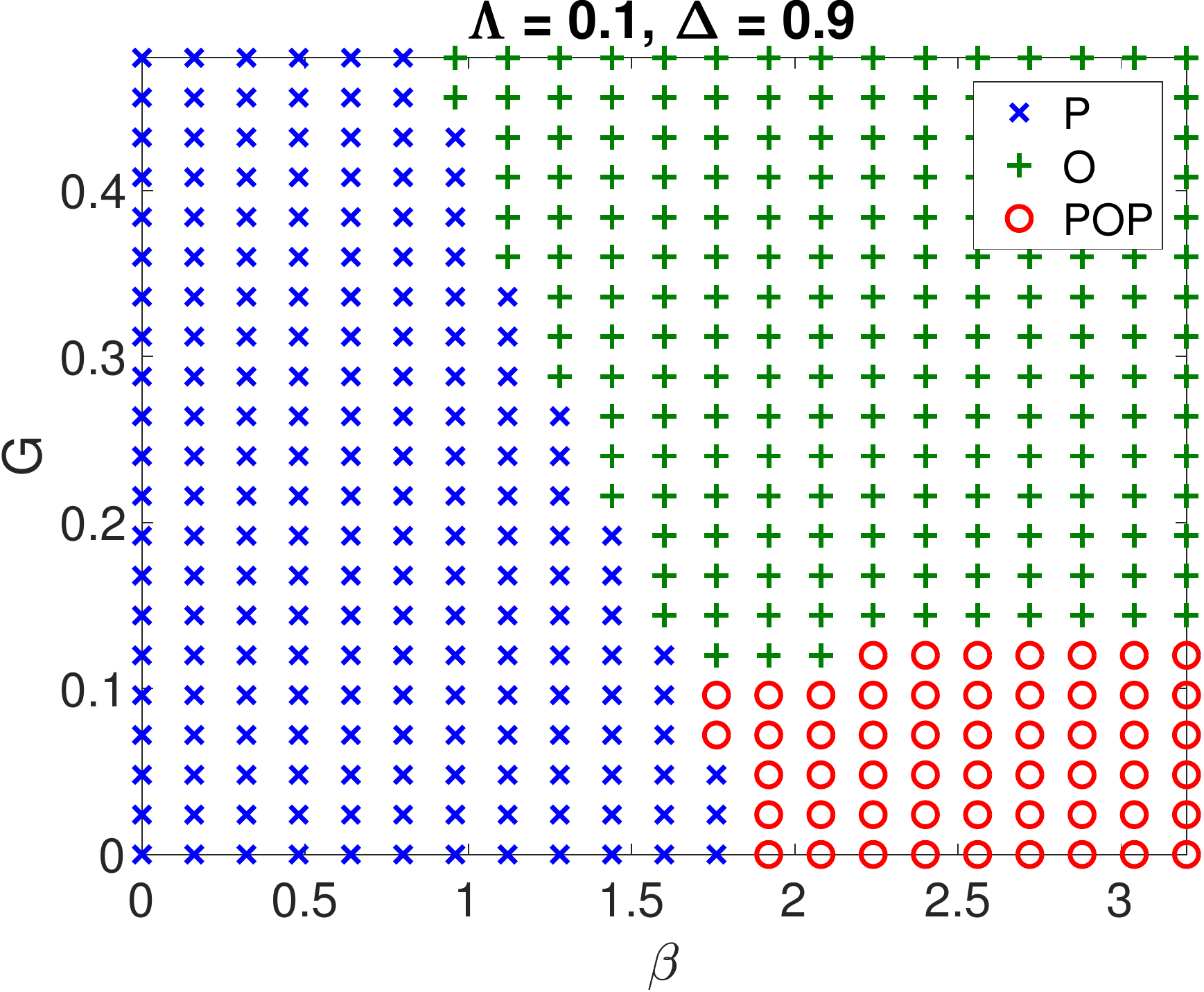}  \includegraphics[height = 1.4in]{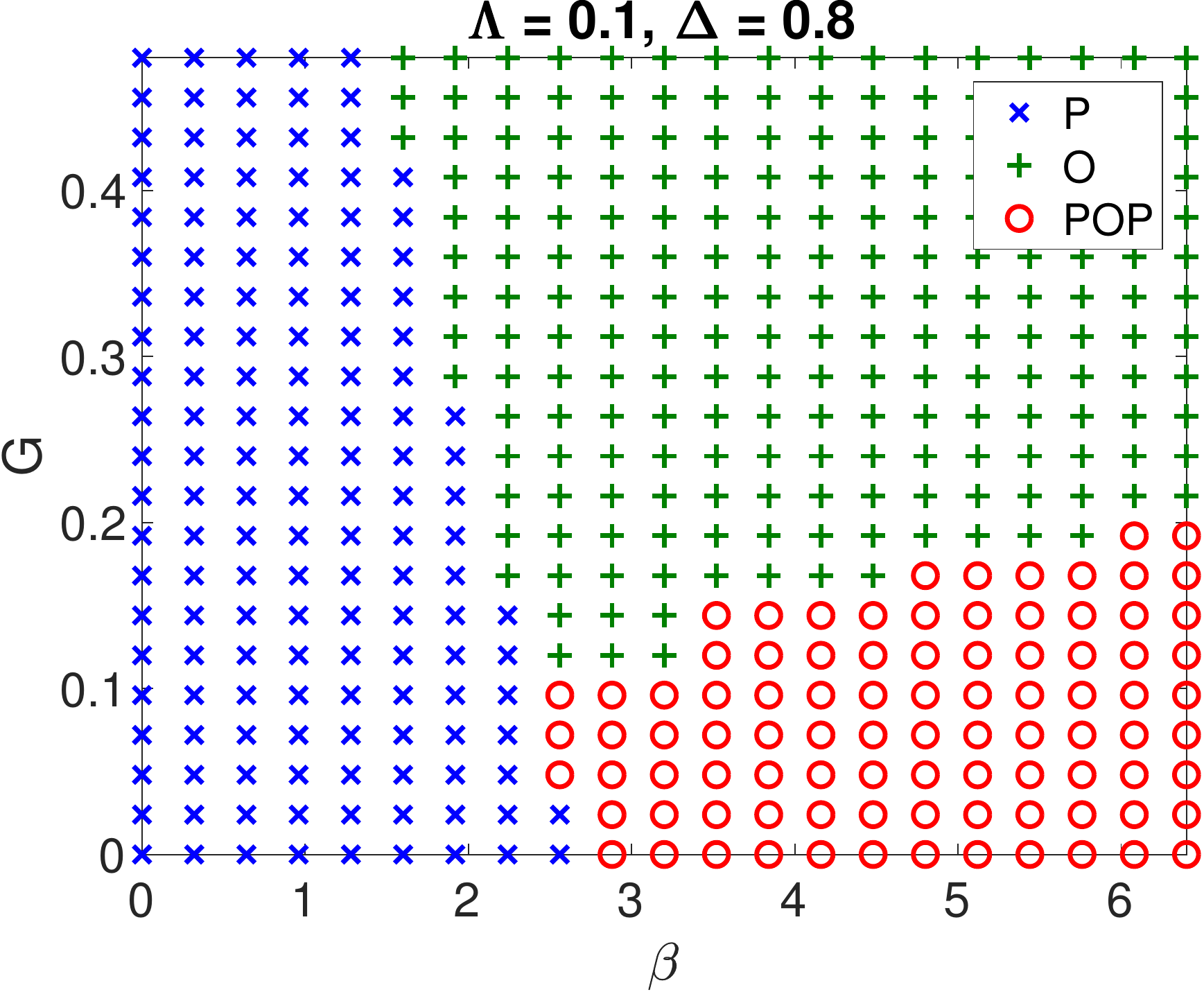}   \includegraphics[height = 1.4in]{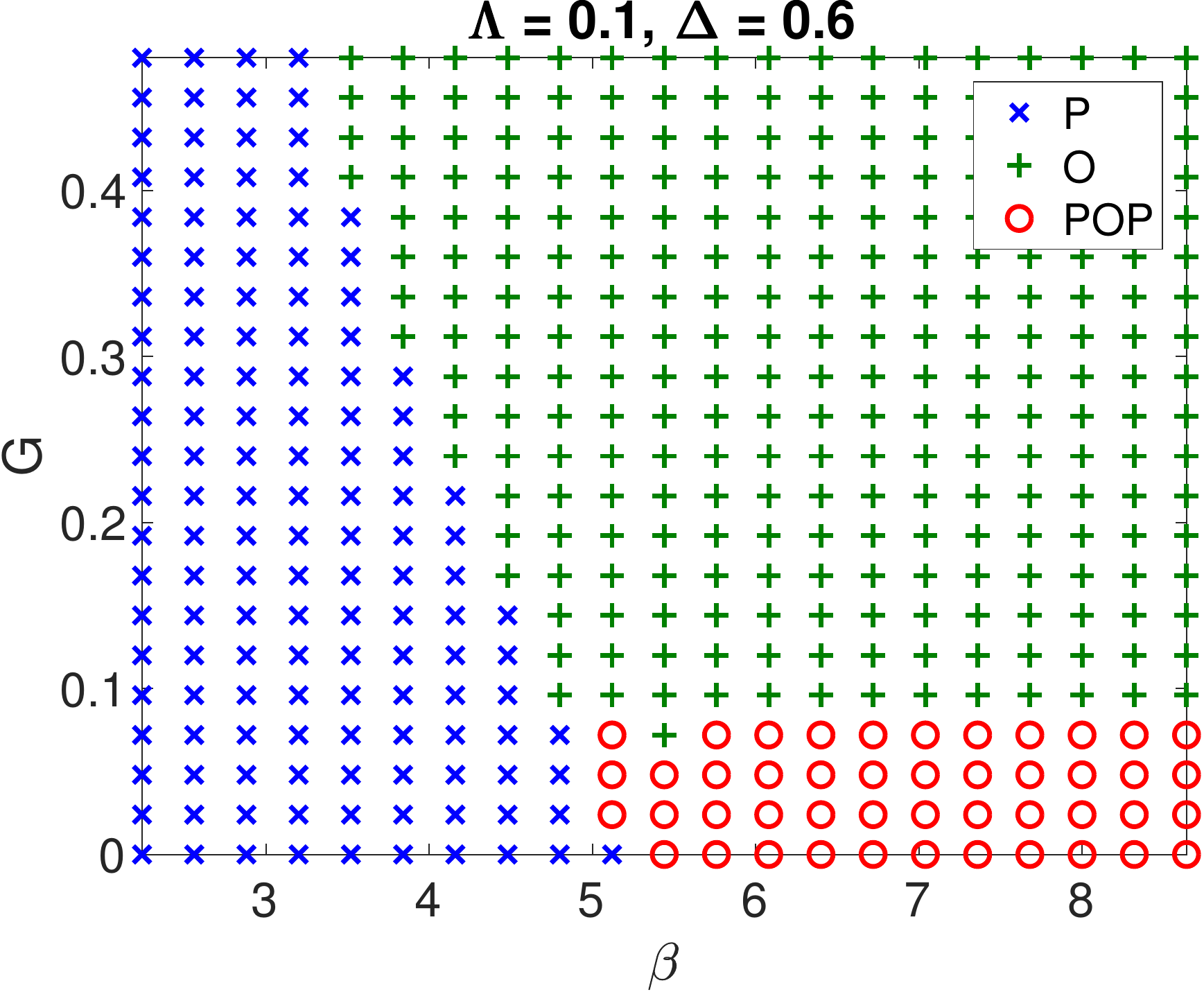} 
\caption{Phase diagrams of vesicle dynamics for different reduced areas as a function of the membrane conductivity $G$ and electric field strength $\beta$. Here, the different phases of the dynamics are indicated by $O$ when the vesicle remains oblate for all times or $P$ when it remains prolate or POP when it transitions from prolate to oblate to prolate shapes. For all the cases, the conductivity ratio $\Lambda$ is set to $ 0.1$, $\mathit{Ca} = 0$.
}
\label{fig:POP_vs_G_nd_E0}
\end{center}
\end{figure}

We further characterize the POP mechanism in Figure \ref{fig:POP_vs_G_nd_E0} for different reduced areas. In all the cases, we observe that there exists some critical field strength $\beta_0$ for POP transition to happen (e.g., from the figure, for $G=0$, $\beta_0 \approx \{1.9, 2.6, 5.1\}$ corresponding to $\Delta = \{0.9, 0.8, 0.6\}$ respectively). On the other hand, when the field strength is weak, the vesicle remains a prolate and when the membrane conductivity is  high, it transitions to an equilibrium oblate shape. These results are in qualitative agreement with \cite{C5SM00585J}, where similar phase diagrams were presented but only for higher reduced area vesicles. Thus the phase diagrams in Figure \ref{fig:POP_vs_G_nd_E0} show that the POP mechanism works consistently for different $\Delta$.

Finally, in the case when $\Lambda > 1$, the EHD forces act to extend the vesicle and it remains a prolate throughout the simulation.

\subsection{Electro-rheology in the dilute limit}\label{sc:one-ves-rheo}
We next look at the combined effect of an imposed shear flow and a DC electric field on a single vesicle. In the presence of both fields, the dynamics is characterized by a competition between the electrical and hydrodynamical shear stresses and the migration of  electric charges along the vesicle membrane. 

\begin{figure}[t]
	\begin{center}
		\includegraphics[height=0.261\textwidth]{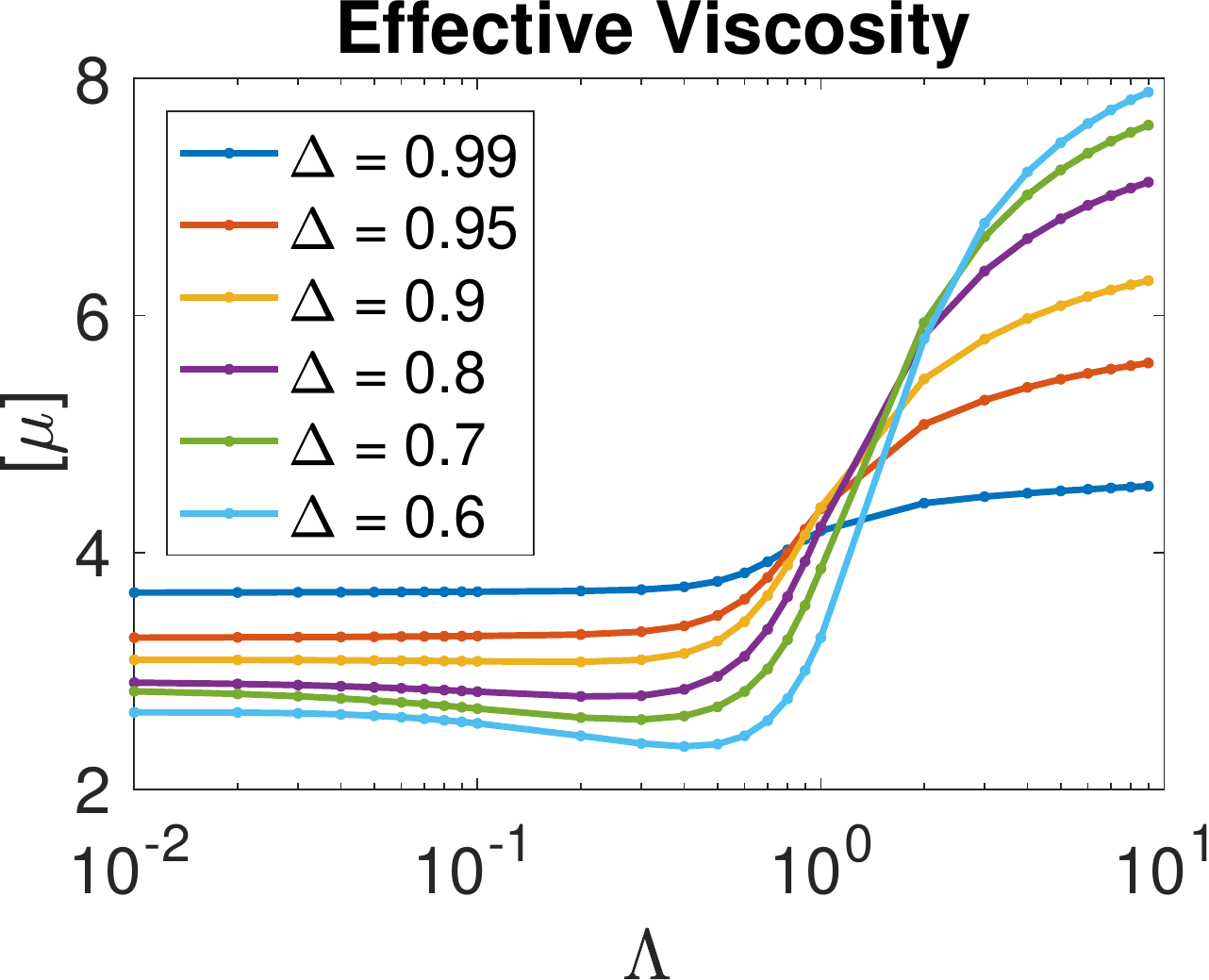}
		\includegraphics[height=0.261\textwidth]{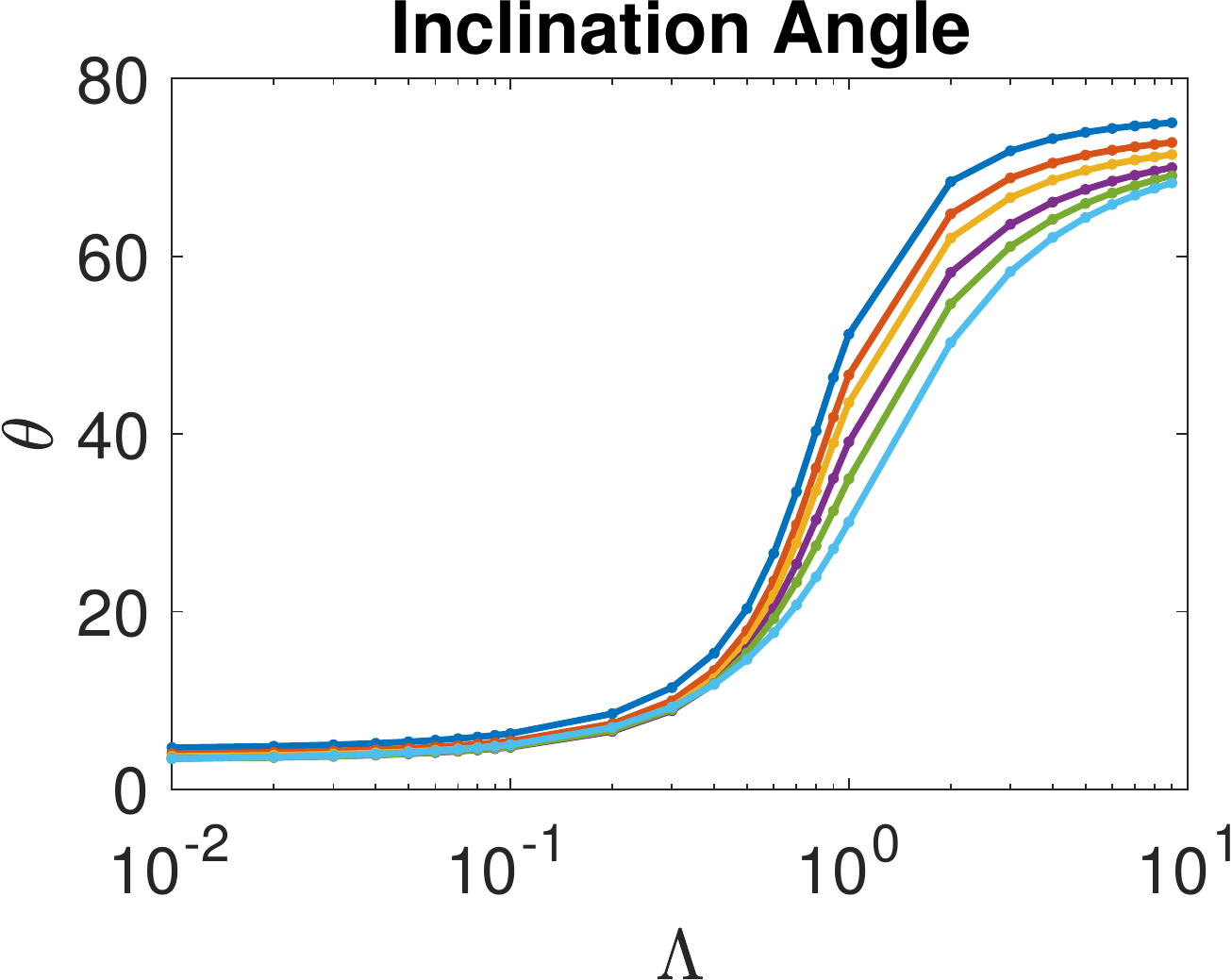}
		\includegraphics[height=0.261\textwidth]{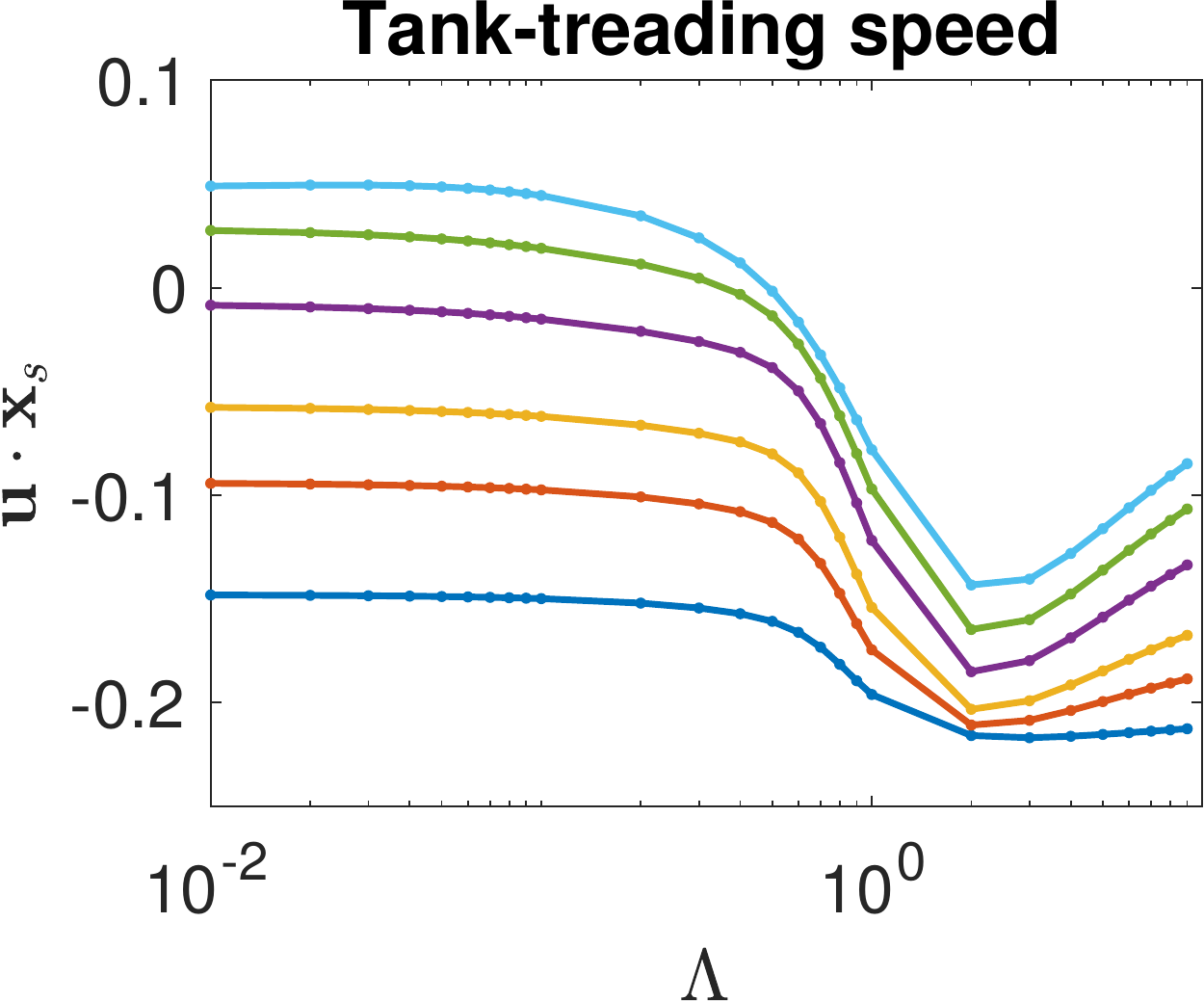}   
		\caption{Single vesicle rheology when $G=4$, $\beta=6.4$ and $\mathit{Ca}=10$. Plots of the effective viscosity (left), angle of inclination (middle) and the tangential velocity (right) when a vesicle is suspended in a linear shear flow as a function of the conductivity ratio. We can observe that the inclination angle increases as $\Lambda$ is increased i.e., the vesicle tries to align with the electric field direction and away from the direction of shear. Thereby, it presents more resistance to imposed flow,  leading to higher effective viscosity. One remarkable effect of low reduced area, as is evident from the right panel, is that the vesicle tank-treads in the opposite direction compared to high reduced area vesicles when $\Lambda$ is small.} 
		\label{fig:rheology}
		\end{center}
\end{figure}

Figure \ref{fig:rheology} shows the rheological properties of a vesicle subjected to an applied linear shear and an applied uniform electric field. In this case, where the membrane has non-zero $G$, we observe that the vesicles with different reduced areas all stabilize into a tank-treading motion and that the tank-treading speed and angle of inclination are affected nonlinearly by the conductivity ratio $\Lambda$. Note that as $\Lambda$ is increased, the vesicle tries to align with the electric field direction and away from the direction of shear, presenting higher resistance to the imposed flow and hence leading to higher effective viscosity. Here, the effective viscosity $[\mu]$ is computed using the usual formula 
\citep{rahimian10}: 
\begin{equation}
[\mu]:= \dfrac{1}{\dot{\gamma} \mu (T_e-T_i)}\int_{T_i}^{T_e} \langle \sigma^p_{12}\rangle dt,
\quad\text{where}\quad
\langle \sigma^p \rangle = \dfrac{1}{A}\int_\gamma (\mbf_b + \mbf_\lambda - \mbf_{el})\otimes \mbx\, ds,
\end{equation}
$A$ is the area of the vesicle and $\sigma^p$ represents the perturbation in the stress due to membrane forces. After the vesicle reaches a steady-state, the effective viscosity is measured over an arbitrary time interval $[T_i, T_e]$.

\begin{figure}[t]
\centering
\includegraphics[width=.95\textwidth]{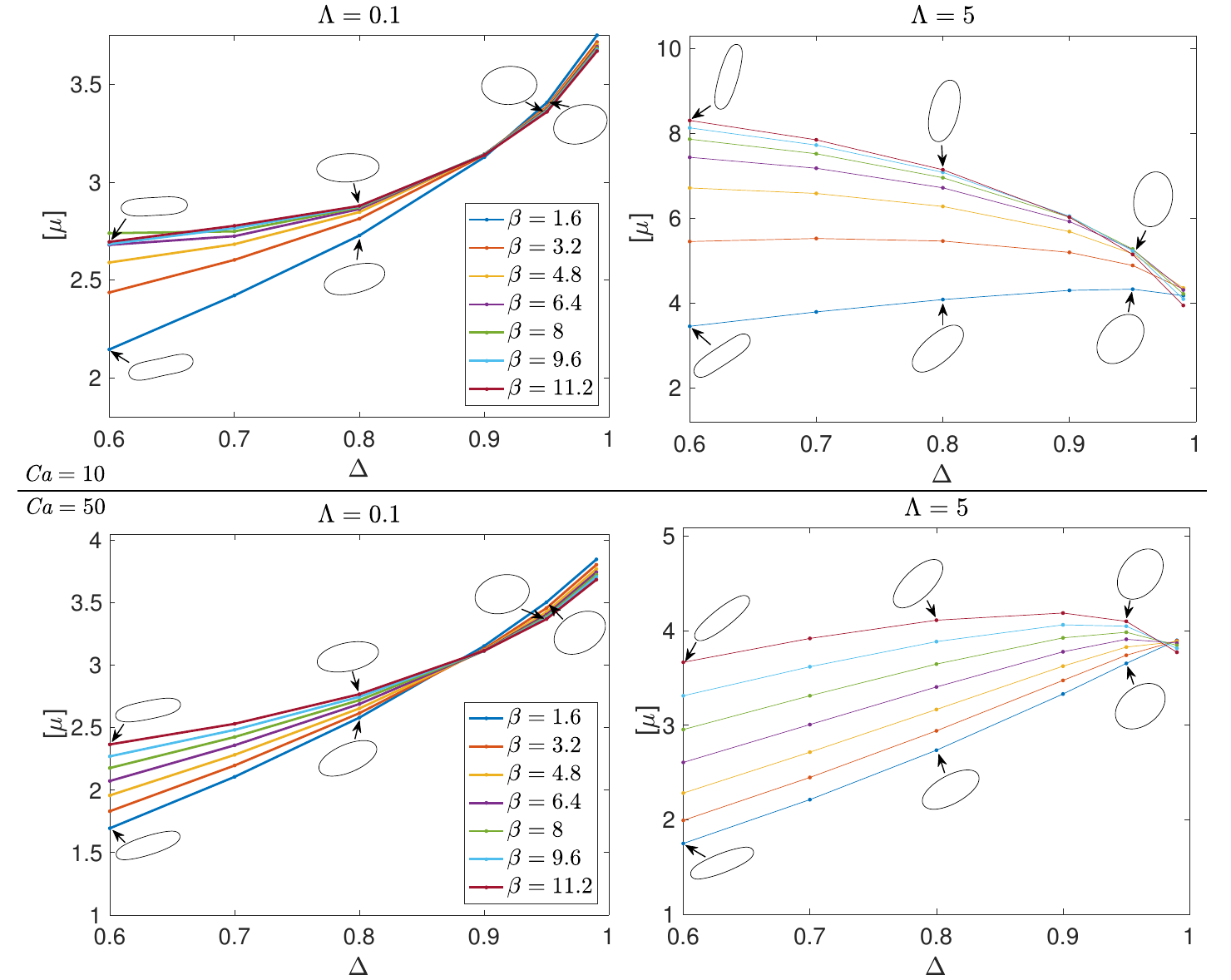}
\caption{Dependence of effective viscosity $[\mu]$ on $\beta$ and $\Delta$. Conductivity $G=4$ and $\mathit{Ca} = 10$ (top row) or $\mathit{Ca} = 50$ (bottom row). We note that (i) $[\mu]$ is higher whenever the equilibrium angle at which the vesicle tank-treads is away from the direction of shear and (ii) when $\Delta$ is close to $1$ (vesicle closer to a circle), $[\mu]$ is nearly $\beta$-independent and shear-independent (as can be expected).}
\label{fig:rheology_betaDelta}
\end{figure}

We further characterize the rheology in Figure \ref{fig:rheology_betaDelta} by plotting the effective viscosity as $\Delta$ is varied. Highly deflated vesicles prominently display shear-rate and $\beta$-dependent rheology since their shapes at equilibrium tank-treading dynamics are different, thereby, presenting varied resistance to applied shear. 


In the case when $G$ is set to zero, the rheological behavior becomes much more complex, primarily because of the tendency of vesicles to undergo a POP transition while at the same time tank-tread due to the applied shear. For different values of $\Lambda$ and $\Delta$, we observed various behaviors such as tumbling, staggering (tank-treading with periodically varying inclination angles), ``mirrored'' tank-treading (tank-treading in the opposite direction and with inclination against the applied shear direction), and even chaotic staggering. A detailed analysis and characterization of these dynamics are beyond the scope of the present work and will be reported at a later date. 

\subsection{Two-body EHD interactions} 

\begin{figure}[t]
\centering
\includegraphics[height=2.5in]{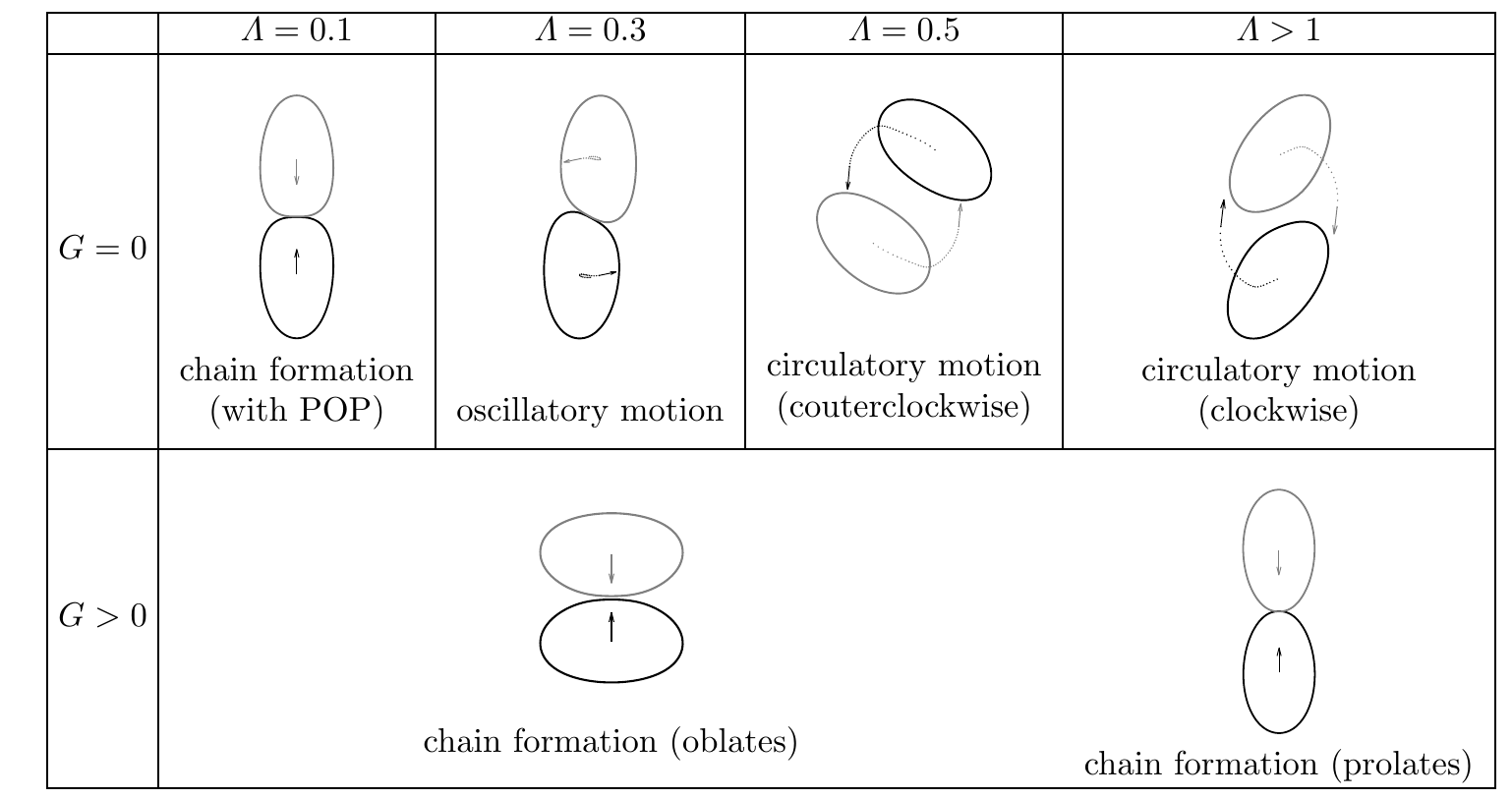}
\caption{A summary of pairwise vesicle EHD interactions ($\Delta = 0.9$, $\beta=3.2$, $\mathit{Ca} = 0$)}
\label{tbl:2ves}
\end{figure}

Next we present results from simulation of two-body vesicle interactions in applied electric field and in the absence of imposed flow. As before, we assume that the viscosity and permittivity of the interior and exterior fluids are the same. We set the initial shape of both the vesicles to be identical and their initial location not symmetric with respect to the electric field direction\footnote{When they are aligned along $\mbe_\infty$, they simply attract each other (after transient shape changes) and when aligned in the perpendicular direction, they simply repel each other---both results are consequences of one vesicle appearing to the other as a dipole with same orientation.}. We apply a DC electric field, pointing upwards, strong enough to cause the POP transition when $\Lambda = 0.1$  (i.e., $\beta > \beta_0$). Under these conditions, the different representative classes of dynamics observed are summarized in Figure \ref{tbl:2ves}. 

The complex nature of these pairwise interactions can be understood from three predominant, competing mechanisms: (i) The electrically-driven vesicle alignment due to one vesicle appearing as a dipole (to leading order) in the far-field electrical disturbance produced by the second vesicle. The two vesicles always tend to form a chain along the direction of dipole orientation; (ii) The EHD flow induced by the tangential electrical stresses at the fluid-vesicle interfaces, driving the vesicles to rotate about each other; (iii) The prolate-oblate deformation mentioned in Section \ref{sc:one-ves-deform}, generating extensional flows around each vesicle. 

First, let us consider the case of $G=0$ i.e., the vesicle membranes are impermeable to charges. Three different types of dynamics can be observed from Figure \ref{tbl:2ves}. The first is {\em chain formation}, observed when $\Lambda$ is small enough, wherein, pronounced deformation, due to mechanism (iii), induces flows that dominate the circulatory flow of mechanism (ii). Thereby, it completely halts the tank-treading motion. At the end of their POP cycle, both vesicles become almost vertically-aligned. Then, mechanism (i) slowly drives them to form a stable chain. From our numerical experiments, we noticed that the thin layer of fluid between the vesicles gets continuously drained albeit at a very slow pace (distance between them decays exponentially with time). 

The second type is a {\em circulatory motion}, observed when $\Lambda$ is large enough, wherein, mechanism (iii) becomes negligible. As the two vesicles move to form a chain, mechanism (ii) causes both of them to tank-tread. Consequently, the induced disturbance flow on each vesicle becomes dominant and they start to rotate about each other. The tank-treading motion also causes the vesicles to appear as tilted dipoles, so they tend to form a tilted chain. The circulatory motion is periodically reinforced by the tilted-chain formation process. The direction of rotation depends on the net torque on each vesicle, which has opposite orientations for $\Lambda > 1$ and $\Lambda \leq 1$.

\begin{figure}[t]
	\begin{center}
		\includegraphics[width=0.8\textwidth]{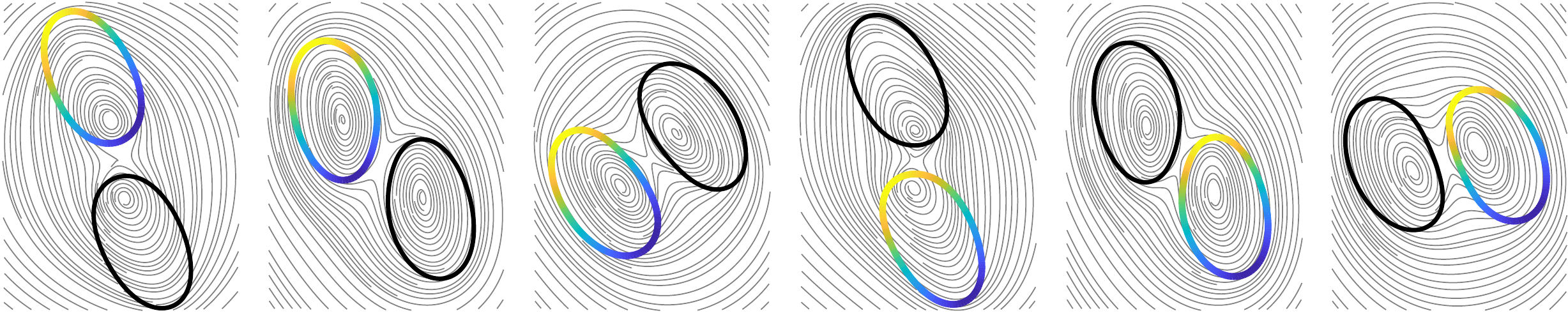}  
		\caption{ Snapshots from a simulation of two vesicles undergoing circulatory motion described in Figure \ref{tbl:2ves} with $G = 0$ and $\Lambda = 0.5$. Here, one of the vesicles is colored by the magnitude of $V_m$ (yellow indicates positive and blue indicates negative values respectively). We can observe that each vesicle undergoes tank-treading motion on its own (as indicated by the streamlines), they rotate about each other and the vesicles viewed as dipoles are always tilted with respect to the applied field direction. }
		\label{fig:circulatory_with_Vm}
		\end{center}
\end{figure}

The last type is an {\em oscillatory motion}, where the two vesicles form an unstable chain and oscillate about each other. This is a transitional situation between the first two types, observed when $\Lambda$ is between the values of those types. In this case, neither the circulatory flow of mechanism (ii) is strong enough to keep vesicles rotating about each other nor the deformational flow of mechanism (iii) is strong enough to completely halt the rotations. The two vesicles tend to form a chain that is periodically tilted one way or the other; each time the vesicles passing a tilted-chain position, tank-treading slows down and the dipole orientation oscillates back. Therefore, mechanisms (i) and (ii) collaborate to keep the vesicles oscillating near the vertical chain position.

On the other hand, the dynamics are much simpler when the membrane is permeable to charges i.e., $G \gg 0$. After a very short period of initial charging, the electric stresses become almost normal to the surface of each vesicle, so mechanism (ii) doesn't arise at all. By mechanism (iii) the vesicles eventually become oblate when $\Lambda < 1$ (with strong enough $\beta$) and become prolate when $\Lambda > 1$, and mechanism (i) drives the vesicles to form a vertical chain. 

\begin{figure}[t]
\includegraphics[width = 0.33\textwidth]{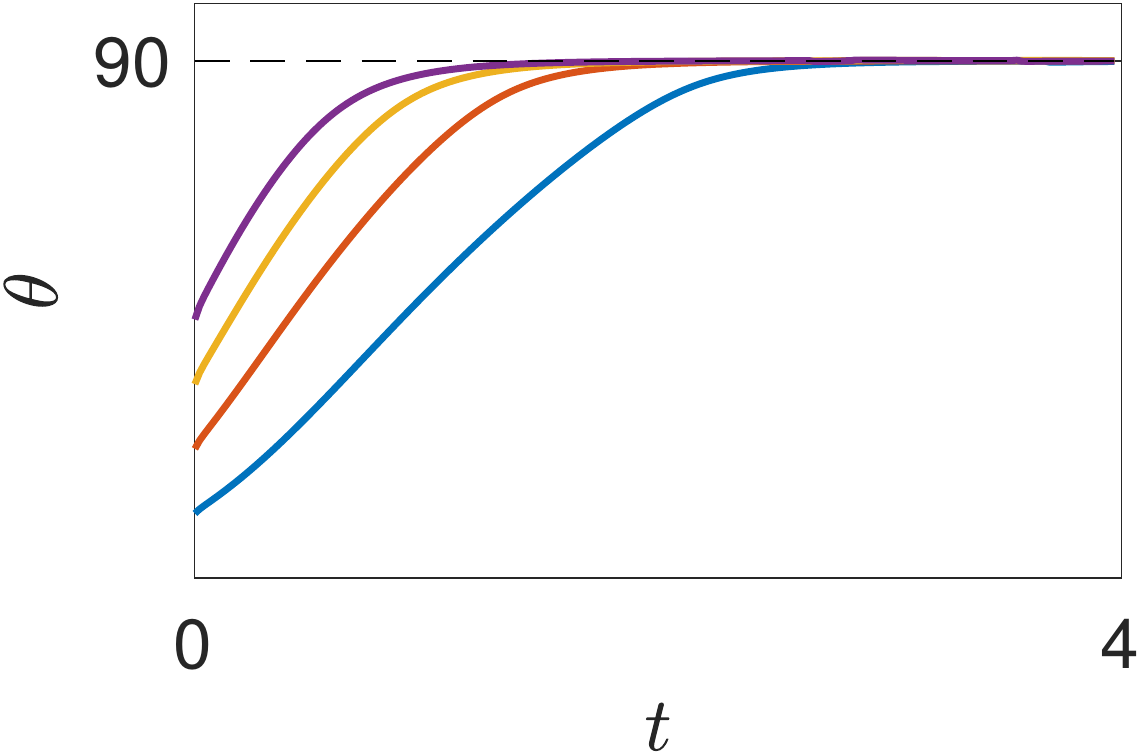}
\includegraphics[width = 0.33\textwidth]{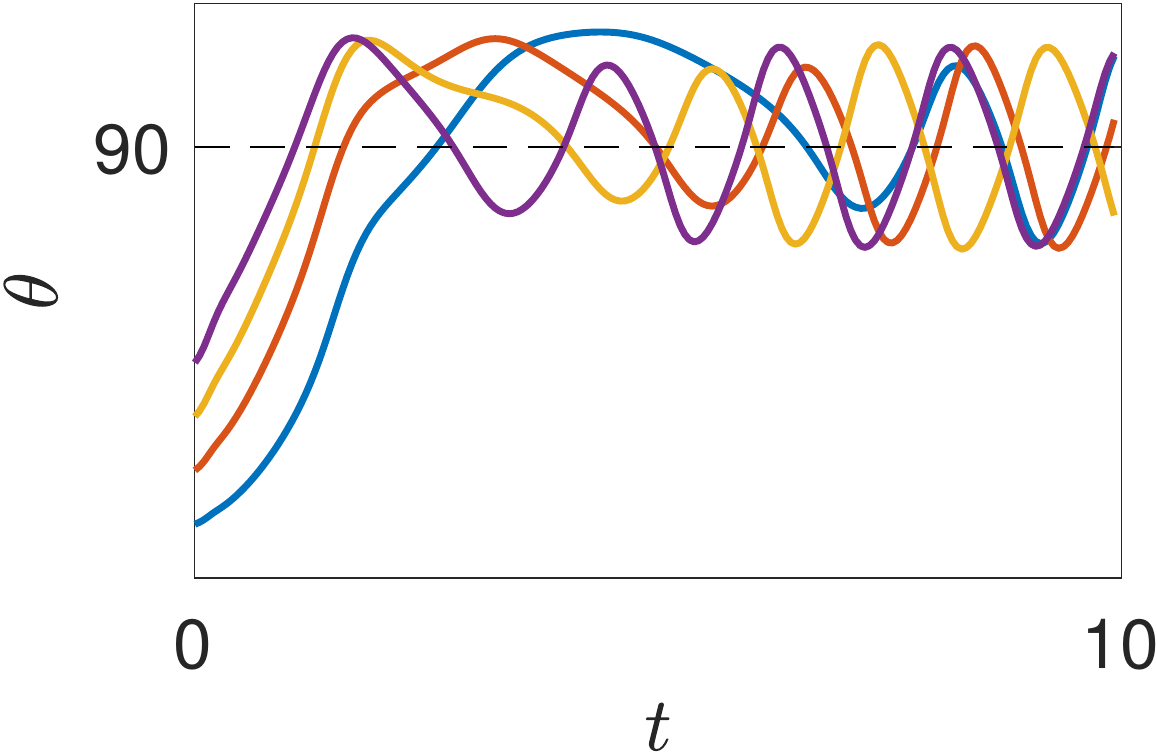} 
\includegraphics[width = 0.33\textwidth]{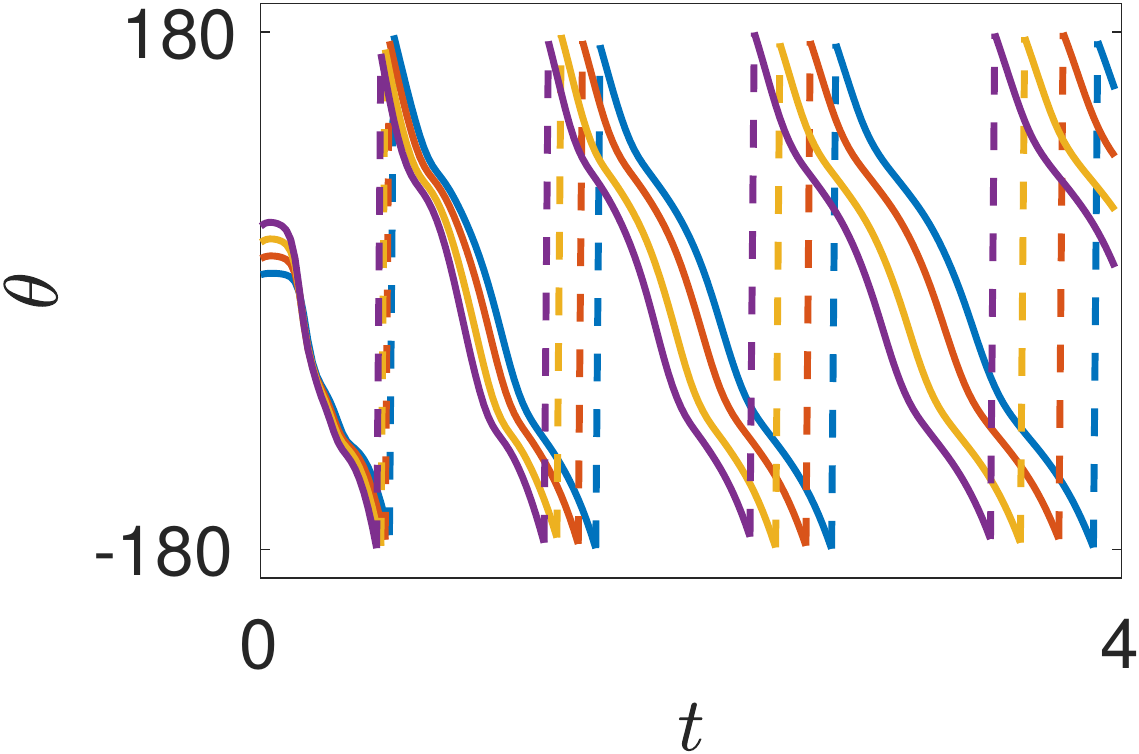}
\caption{Insensitivity of the EHD pairwise interactions to the initial offset from the aligned position. $\theta$ measures the angular offset of the two vesicles relative to the horizontally aligned position. (a) Chain formation. (b) Oscillatory motion. (c) Circulatory motion. In each case, the same pattern is observed regardless of the initial $\theta>0$.} 
\label{fig:insensitive_angle}
\end{figure}

\begin{figure}[t]
\centering
\includegraphics[width=.99\textwidth]{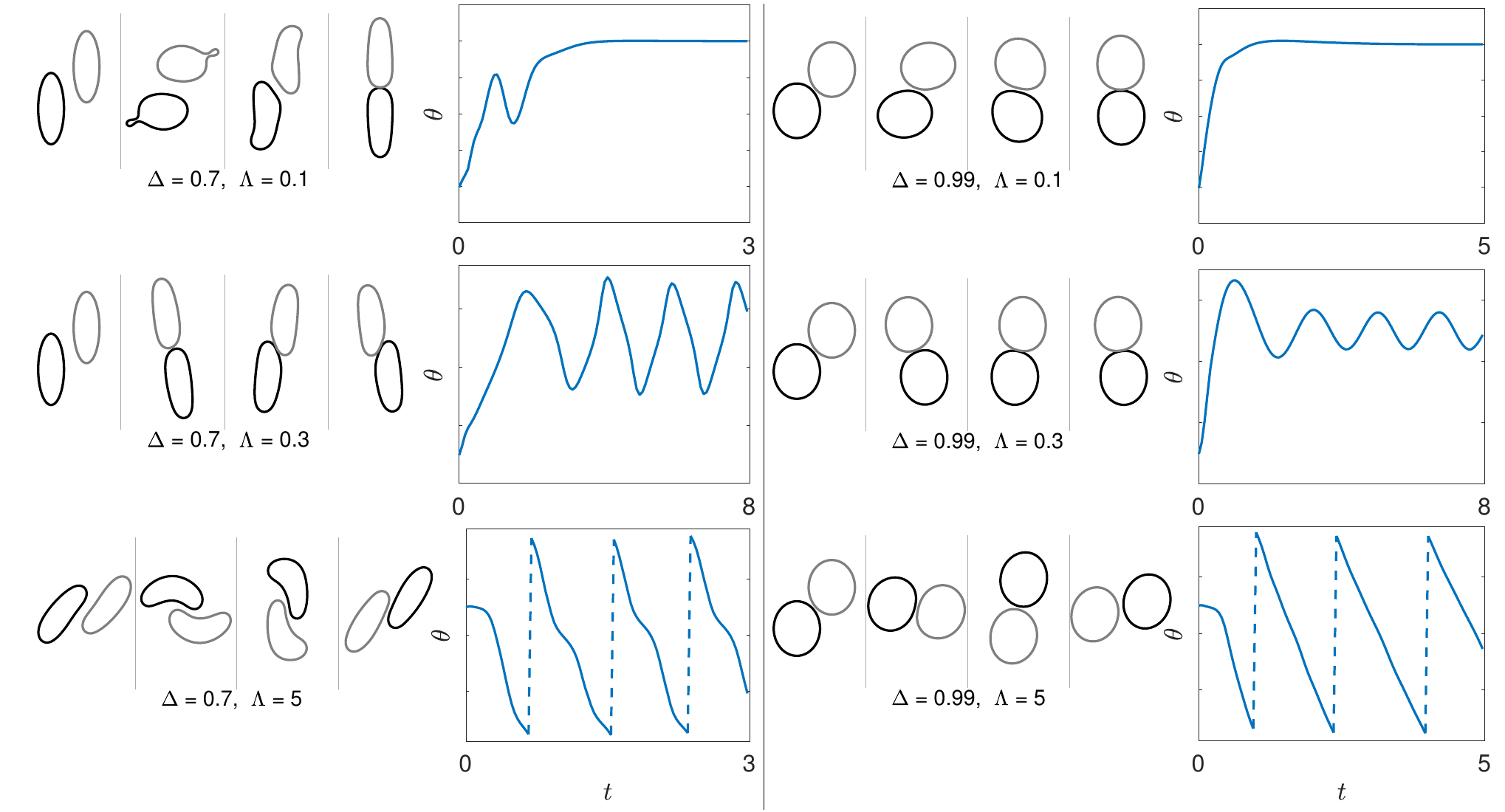}
\caption{Pairwise interactions for $G=0$ vesicles of reduced areas $\Delta = 0.7$ (with $\beta = 4.8$) and $\Delta = 0.99$ (with $\beta = 2.4$). $\mathit{Ca} = 0$. The behaviors (e.g. chain formation, oscillatory motion, circulatory motion) are the same as in the $\Delta = 0.9$ case (Fig. \ref{tbl:2ves}), showing that the mechanism of pairwise interactions is insensitive to the reduced area. Note that the bud formation also happens with the case of $\Delta=0.7, \Lambda=0.1$.}
\label{fig:pairwise_varyDelta}
\end{figure}

\textbf{Sensitivity to positions and shapes.} Note that all of the aforementioned dynamics are insensitive to the initial offset or shapes of the vesicles. In Figure \ref{fig:insensitive_angle}, we demonstrate that for different initial angular offsets from the aligned position, the vesicles undergo the same type of pairwise interaction that corresponds to the given $\Lambda$ and $G$. Furthermore, Figure \ref{fig:pairwise_varyDelta} shows that the similar kind of dynamics are observed for vesicles with different reduced areas, therefore, the pairwise EHD interaction mechanisms appear to be consistent for highly-deflated or close-to-circular vesicles.

\begin{figure}[t]
\centering
\includegraphics[width=\textwidth]{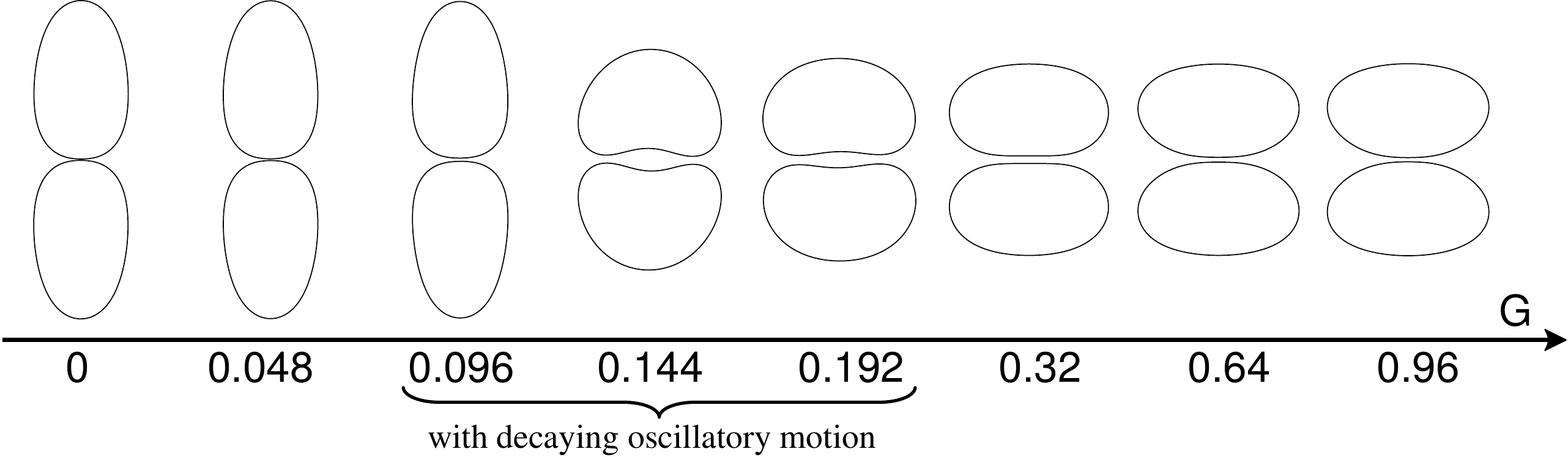}
\includegraphics[height=.3\textwidth]{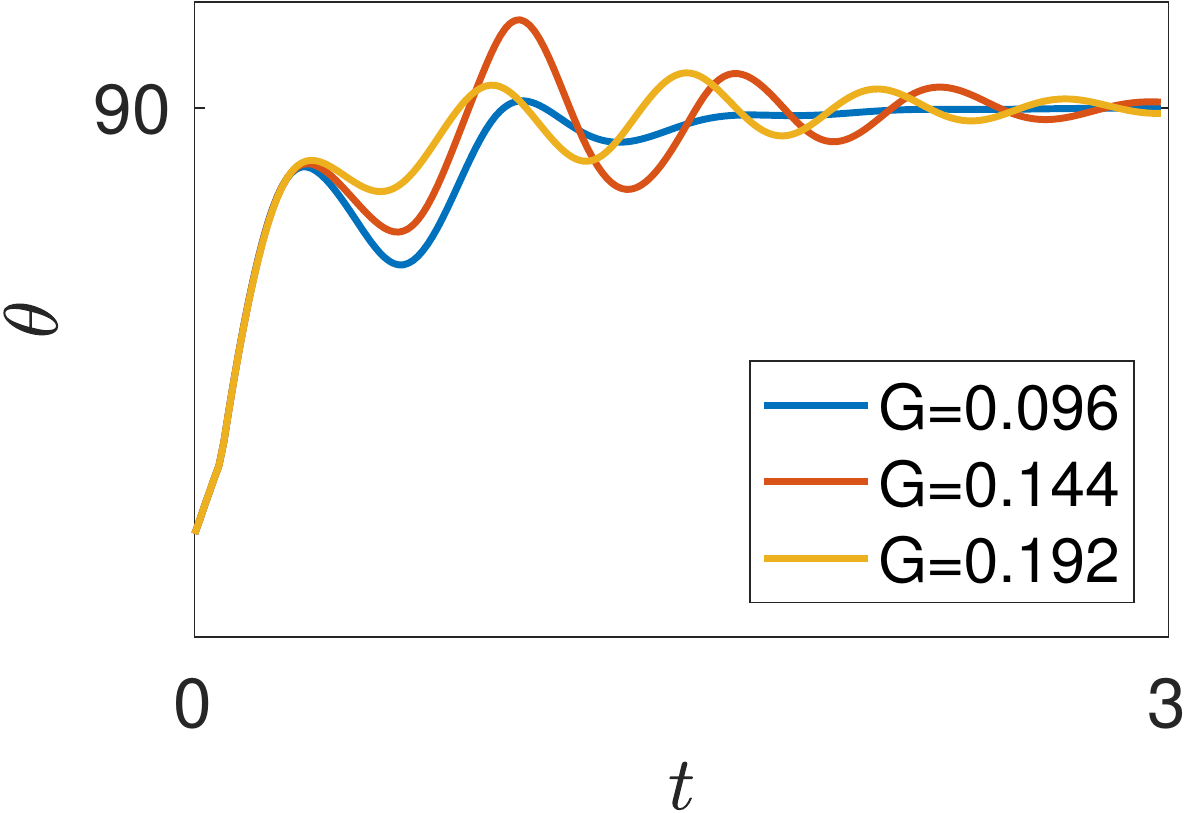}\qquad \includegraphics[height=.3\textwidth]{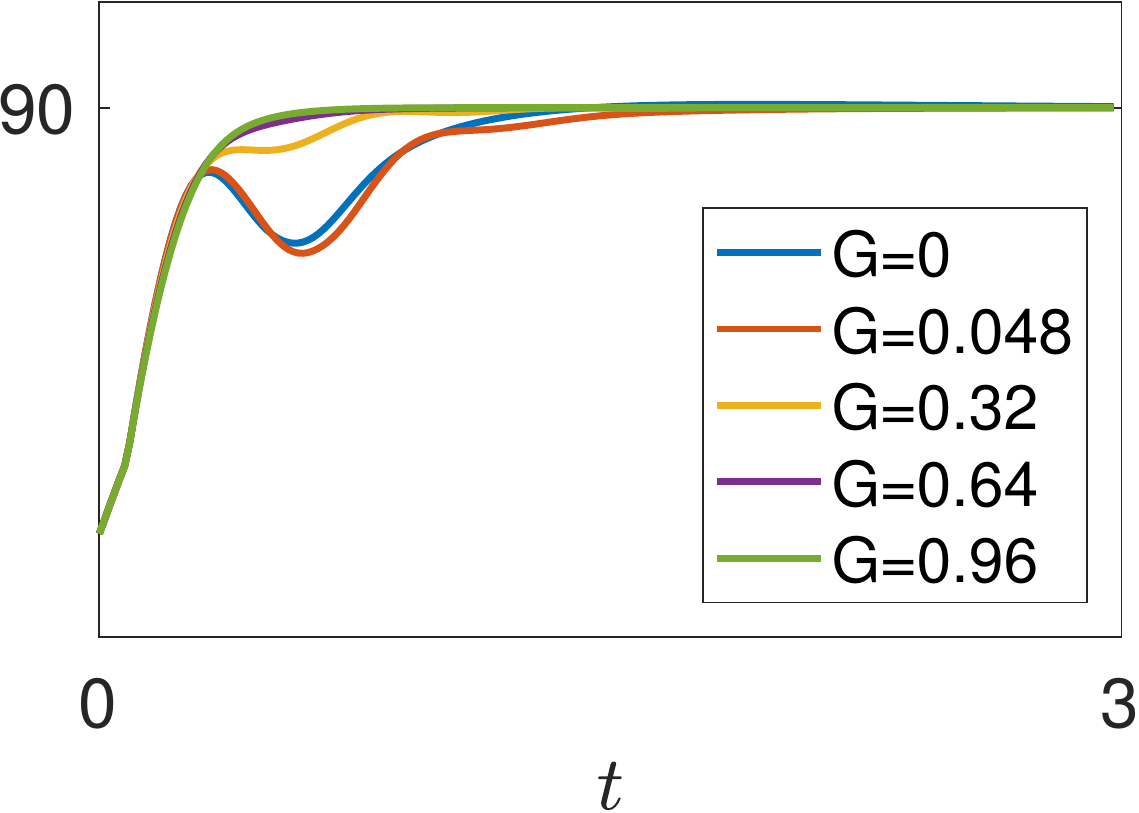}\qquad
\caption{Top figure: final configurations 
of eight separate simulations, each corresponding to a different membrane conductivity $G$. There is a continuous transition from a chain of prolates ($G\approx0$) to a chain of oblates ($G\gg0$). For certain intermediate value of $G$ (e.g. $G = 0.096, 0.144, 0.192$) the chain formation process is accompanied with decaying oscillatory motions (lower left figure), while for more extreme values of $G$ the vesicles directly form a chain without oscillations (lower right figure). For all simulations $\beta = 3.2$, $\Lambda = 0.9$, and $\mathit{Ca} = 0$.}
\label{fig:pairwise_intermediate_G}
\end{figure}

\textbf{Continuous transition.} Finally, we note that the dynamics transitioning from $G=0$ to $G>0$, as shown in Figure \ref{tbl:2ves}, are not abrupt. To illustrate this, we show in Figure \ref{fig:pairwise_intermediate_G} the pairwise dynamics of vesicles with $\Lambda=0.1$, demonstrating a continuous transition from a chain of prolates ($G=0$) to a chain of oblates ($G\gg0$); for certain intermediate values of $G$, one can even observe interesting kidney-like shapes as well as decaying oscillations of the vesicles as they settle into their equilibrium shapes.

\section{Conclusions}
We investigated the pairwise dynamics and rheology of vesicles in DC electric fields using a boundary integral method. Our method is shown to reproduce previous results on isolated vesicle EHD and can be extended in a trivial manner to study the EHD of large number of vesicles. We showed that much richer set of pairwise interactions can be observed when the membranes are impermeable to charges. This is somewhat unique to vesicle EHD compared to other systems such as drops \citep{baygents1998electrohydrodynamic}, driven mainly by the capacitative nature of the membranes. However, we explored only a small fraction of the possible dynamics; relaxing our simplifying assumptions---varying the viscosity and permittivity contrasts, imposing an AC electric field, accounting for charge convection along the membrane---is expected to enrich the space much further.  We are currently exploring these as well as analyzing the collective dynamics of dense suspensions in periodic domains. Another important direction we are currently pursuing is to extend our numerical scheme to handle more general EHD models such as those discussed in the recent work of \cite{mori_young_2018}.  
%
%
%
%
\bibliography{3dVesicles,petia}		
\bibliographystyle{plainnat}
\end{document}